\documentclass[lettersize,journal]{IEEEtran}
\usepackage{amsmath,amssymb,amsfonts}
\usepackage{algorithmic}
\usepackage{graphicx}
\usepackage{textcomp}
\usepackage{wrapfig,colortbl}
\usepackage{array}
\usepackage[caption=false,font=normalsize,labelfont=sf,textfont=sf]{subfig}
\usepackage{textcomp}
\usepackage{stfloats}
\usepackage{url}
\usepackage{verbatim}
\usepackage{graphicx}
\usepackage{cite}
\usepackage{balance}
\usepackage{breqn}
\usepackage[font=footnotesize,skip=2pt]{caption}
\usepackage[caption=false]{subfig}
\usepackage[linesnumbered,ruled,vlined]{algorithm2e}

\SetCommentSty{mycommfont}
\def\BibTeX{{\rm B\kern-.05em{\sc i\kern-.025em b}\kern-.08em
    T\kern-.1667em\lower.7ex\hbox{E}\kern-.125emX}}

\begin{document}

\title{CapsBeam: Accelerating \textbf{Caps}ule Network based \textbf{Beam}former for Ultrasound Non-Steered Plane Wave Imaging on Field Programmable Gate Array}

\author{Abdul Rahoof,~\IEEEmembership{Student Member, IEEE}, Vivek Chaturvedi,~\IEEEmembership{Senior Member, IEEE}, Mahesh Raveendranatha Panicker,~\IEEEmembership{Senior Member, IEEE}, and Muhammad Shafique,~\IEEEmembership{Senior Member, IEEE}
\thanks{Abdul Rahoof and Vivek Chaturvedi are with the Indian Institute of Technolgy Palakkad (IITPKD), India (e-mail: 112004001@smail.iitpkd.ac.in; vivek@iitpkd.ac.in). }
\thanks{Mahesh Raveendranatha Panicker is with the Singapore Institute of Technology (SIT), Singapore (e-mail: mahesh.panicker@singaporetech.edu.sg ).}
\thanks{Muhammad Shafique is with the New York University Abu Dhabi, UAE (e-mail: muhammad.shafique@nyu.edu).}}

\markboth{IEEE TRANSACTIONS ON VERY LARGE SCALE INTEGRATION (VLSI) SYSTEMS,~Vol.~14, No.~8, August~2021}%
{Abdul Rahoof \MakeLowercase{\textit{et al.}}: CapsBeam: Accelerating \textbf{Caps}ule Network based \textbf{Beam}former for Ultrasound Non-Steered Plane Wave Imaging on Field Programmable Gate Array}


\maketitle

\begin{abstract}
In recent years, there has been a growing trend in accelerating computationally complex non-real-time beamforming algorithms in ultrasound imaging using deep learning models. However, due to the large size and complexity these state-of-the-art deep learning techniques poses significant challenges when deploying on resource-constrained edge devices. In this work, we propose a novel capsule network based beamformer called CapsBeam, designed to operate on raw radio-frequency data and provide an envelope of beamformed data through non-steered plane wave insonification. Experiments on in-vivo data, CapsBeam reduced artifacts compared to the standard Delay-and-Sum (DAS) beamforming. For in-vitro data, CapsBeam demonstrated a $32.31\%$ increase in contrast, along with gains of $16.54\%$ and $6.7\%$ in axial and lateral resolution compared to the DAS. Similarly, in-silico data showed a $26\%$ enhancement in contrast, along with improvements of $13.6\%$ and $21.5\%$ in axial and lateral resolution, respectively, compared to the DAS. To reduce the parameter redundancy and enhance the computational efficiency, we pruned the model using our multi-layer LookAhead Kernel Pruning (LAKP-ML) methodology, achieving a compression ratio of $85\%$ without affecting the image quality. Additionally, the hardware complexity of the proposed model is reduced by applying quantization, simplification of non-linear operations, and parallelizing operations. Finally, we proposed a specialized accelerator architecture for the pruned and optimized CapsBeam model, implemented on a Xilinx ZU7EV FPGA. The proposed accelerator achieved a throughput of 30 GOPS for the convolution operation and 17.4 GOPS for the dynamic routing operation. 
\end{abstract}


\section{Introduction}
\par Plane-Wave Imaging~\cite{ultra,pw} is a high frame rate ultrasound imaging technique, wherein the entire medium undergoes insonification through the simultaneous activation of all transducer elements. Consequently, all piezoelectric elements capture the backscattered signals from the medium, resulting in frame rates reaching several thousands of frames per second. However, the resulting image quality is compromised when employing the standard Delay-and-Sum (DAS)~\cite{das} beamforming technique for image reconstruction. To address this, Coherent Plane-Wave Compounding~\cite{cpw} presents an effective solution by combining the beamformed outputs of multiple insonifications. Nonetheless, this approach comes with the drawback of reduced frame rate, leading to a trade-off between image quality and frame rate. Adaptive beamforming techniques such as Minimum Variance Distortionless Response (MVDR)~\cite{mvdr} provide an alternative by using data-dependent apodization weights on the delay-compensated signals. However, the computational demands of MVDR, especially the matrix inversions, presents challenges for real-time implementation, with a complexity of $O(n^3)$, where $n$ represents the aperture size of the transducer array. For instance, it necessitates approximately 98.78 GOPs/Frame for a frame size of (368, 128)~\cite{mvdr2}.
\par The impressive progressions in deep learning motivated researchers to explore its potential applications in ultrasound beamforming~\cite{fcnn, cnn, unet,googlenet}. While traditional methods like MVDR improve image quality, they are computationally expensive. Recent advances in deep learning have shown promising results in reducing artifacts and improving resolution, yet they often demand substantial computational resources, which limits their use in embedded systems. The complexity of these models, such as~\cite{unet} (90 GOPs/Frame) and~\cite{googlenet} (95 GOPs/Frame), makes them suitable for high-performance CPUs or GPUs but presents challenges for implementation on resource-constrained devices. While existing beamforming techniques~\cite{mvdr,unet,googlenet} offer high-quality images, they fail to meet the real-time processing requirements for high-frame-rate ultrasound imaging. The gap addressed in this paper is the need for a computationally efficient, high quality high frame rate ultrasound beamforming technique that can be deployed on resource-constrained edge devices.

\par High frame ultrasound imaging is getting popular due to its applications in analyzing transient motions through B-mode videos as well tissue and blood velocity imaging. Incorporating high frame rate imaging into point of care ultrasound devices is an unsolved challenge and is attempted in this work through a light weight neural network based approach. For high frame rate ultrasound imaging, non-steered (i.e. single angle 0 deg) transmission is employed which will typically result in low quality images. Specialized hardware accelerators provide an efficient solution for handheld ultrasound imaging, offering a superior power-performance tradeoff compared to general-purpose processors in resource-constrained environments. Among these, field-programmable gate arrays (FPGAs) are inherently essential for beamforming control in conventional ultrasound signal chains, and in this work, we leverage them to accelerate beamforming using neural networks. Additionally, transferring data from an FPGA (which is already part of the ultrasound signal chain) to GPU requires a CPU in between and will reduce the frame rate drastically, particularly in the case of plane wave imaging due to high frame rate transmission. Moreover, FPGAs have emerged as a promising choice for beamforming due to their power efficiency, flexibility, and parallel processing capabilities. Their reprogrammability enables adaptable functionality and circuit configurations, making them widely adopted in edge devices for deep learning inference and medical imaging applications~\cite{compact,angel,yolo,cnn-fpga,float,sparse}.
\par Combining the adaptive beamforming acceleration enabled by deep learning and reconfiguration and parallel processing capability of FPGAs, we propose a novel capsule network~\cite{caps} based beamformer called CapsBeam, designed for non-steered insonification plane wave imaging. The capsule networks have recently gained significant traction and demonstrated superior accuracy in various computer vision tasks. Due to its ability to preserve spatial relationships between features, the capsule network is well-suited for data-adaptive beamforming, where maintaining the spatial relationships between signal components is critical. We utilized convolutional capsule layers, as described in~\cite{deepcaps}, with modifications to facilitate easier feature learning and image reconstruction. The calculation pattern used in the dynamic routing algorithm within the capsule layer is distinct from other operations in deep learning models, necessitating a specialized accelerator for the capsule network. We utilized the FastCaps proposed in~\cite{fastcaps} approach for deploying the proposed CapsBeam model on FPGA. To adapt the model on FPGA with limited resources, we compressed the model using a novel multi-layer LookAhead Kernel Pruning (LAKP-ML) method. Additionally, we streamlined the dynamic routing algorithm by simplifying its non-linear operations. Finally, we developed a specialized accelerator architecture for our pruned and optimized CapsBeam model and deployed it on the Zynq UltraScale+ MPSoC ZCU104 FPGA.
\par
\begin{figure}[!t]
\centering
\includegraphics[scale=0.46]{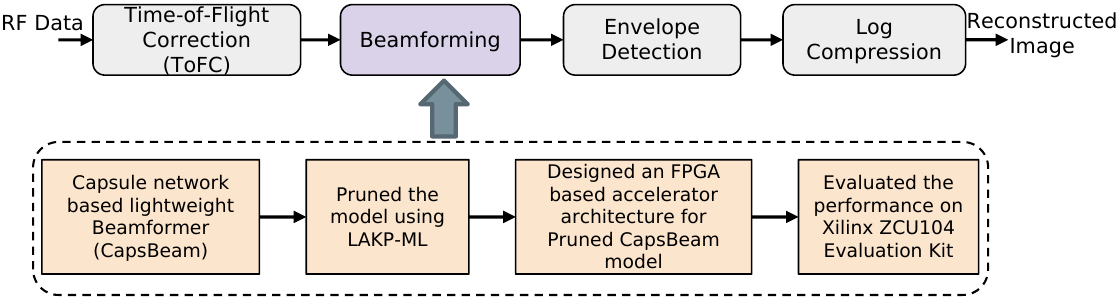}
\caption{Overview of the proposed CapsBeam framework}
\label{fig:my_label-0}
\end{figure}

 \textbf{The main contributions in this work are:} 
\begin{itemize}
    \item We propose a capsule network based ultrasound imaging model, CapsBeam, which improves the quality of reconstructed images in terms of contrast and resolution as compared to the standard DAS~\cite{das} and state-of-the-art tinyML models~\cite{fcnn,cnn,tinyvbf}. Moreover, our CapsBeam is computationally light and uses fewer FLOPs (28.79 GOPs/Frame) compared to the MVDR~\cite{mvdr} beamformer (98.78 GOPs/Frame). 
    \item We extended the LookAhead Kernel Pruning (LAKP) method described in~\cite{fastcaps} by computing the pruning score with weights of more than one succeeding and preceding layers. This extension resulted in an improvement of nearly $0-2\%$ in terms of accuracy on ResNet18 and VGG19, respectively, on CIFAR-10 data. 
    \item The proposed approach pruned the CapsBeam model based on resolution and contrast, focusing on image quality rather than minimizing mean square error, using the novel multi-layer LookAhead Kernel Pruning (LAKP-ML) method and achieved a compression ratio of $85\%$ without compromising the image quality. Moreover, the proposed framework optimized dynamic routing algorithm by simplifying its non-linear operations, reordering the loops, and enabling parallel operations. These optimizations increased the throughput of the dynamic routing operation from 0.37 GOPS to 17.4 GOPS.
    \item  Finally, we quanitized the parameters and designed a specialized accelerator architecture for our pruned and optimized CapsBeam model and deployed it on Xilinx ZCU104 evaluation board with clock frequency 100MHz. The proposed accelerator achieved a throughput of 30 GOPS for the convolution operation and 17.4 GOPS for the dynamic routing operation.
\end{itemize}
To the best of our knowledge, the proposed study represents the first attempt to assess the feasibility of beamforming using a capsule network and deploy the complete model on resource-constrained edge devices. Fig.~\ref{fig:my_label-0} provides an overview of the proposed CapsBeam framework.

\section{Background and Related Works}
\par This section provides a concise overview of different beamforming techniques for plane wave imaging and their associated limitations. Following this, the capsule network is introduced. Finally, network pruning is described as a method to reduce computational complexity within the model.
\subsection{Beamforming Techniques}
\par In ultrasound imaging, the computational tasks can be broadly classified into delay estimation (or time-of-flight correction) and beamforming. Delay estimation can be performed using time domain interpolation, where complexity can be minimized by precomputing delays based on the probe geometry and imaging parameters. On the other hand, beam summation of the delay compensated data can be adaptive based on the tissue reflected signals and hence is computationally intensive. Furthermore, the performance and quality of the reconstructed image are significantly impacted by the beamforming algorithm applied. The conventional Delay-and-Sum (DAS)~\cite{das} is a simple beamformer that utilizes the predefined geometry driven apodization weights to compute the beamformed output. DAS is computationally efficient but provides reconstructed images of poor quality when used with plane wave imaging. Minimum Variance Distortionless Response (MVDR)~\cite{mvdr} beamforming aims at minimizing the variance of the received signal therby having a more directed receive beam. While MVDR produces high-quality images, its computational complexity is high, making it non-trivial to deploy in real-time applications. 
\par Numerous machine learning and deep learning approaches exist to improve a wide range of tasks in signal processing for ultrasound imaging, with a specific emphasis on accelerating beamforming techniques. The work in~\cite{fcnn} devised a fully connected neural network refered to as FCNN for pixel-by-pixel beamforming. This approach exhibited enhanced resolution and contrast compared to the Delay-and-Sum (DAS). However, the FCNN model solely captures local features in the input data to determine apodization weights. The Tiny-CNN beamformer~\cite{cnn} processes data from a region and generates a beamformed image, rather than processing pixel by pixel. This approach improved resolution and contrast with fewer angles of transmission, compared to DAS. However, the quality of images generated through non-steered plane-wave imaging is limited to the performance attained by the DAS method. The works in~\cite{unet,googlenet} introduced extensive convolutional neural networks (CNNs) for image reconstruction. Notably, these approaches significantly improved resolution and contrast in both in-silico and in-vitro data compared to the DAS method, demonstrating notable reductions in sidelobes and artifacts. However, these methods~\cite{unet,googlenet} primarily targeted software-based applications utilizing GPU accelerators and exhibited high computational complexity, exceeding 90 GOPs/Frame for a frame size of (368, 128). Additionally, its implementation on resource-constrained edge devices presents challenges. To the best of our knowledge, there have been limited attempts to develop beamforming accelerators using neural networks for deployment on edge devices. In a recent study~\cite{tinyvbf}, a vision transformer~\cite{attention} based beamformer called Tiny-VBF was introduced, which demonstrated better resolution and contrast on in-silico and in-vitro data compared to DAS and Tiny-CNN. Additionally, the model was deployed on an FPGA platform. However, the model lacks flexibility in handling varying frame sizes, as it supports only a depth of 368 rows for inference. In this paper, we propose a capsule network-based beamformer for non-steered plane wave imaging that achieved better image quality compared to the DAS, Tiny-CNN and Tiny-VBF on in-vivo, in-vitro, and in-silico data.

\subsection{Capsule Network}
\par The capsule network, introduced by~\cite{caps}, enhances the model's capability to retain spatial features and enhances its learning capacity compared to CNNs. The study implemented a fully connected dynamic routing algorithm, demonstrating superior accuracy on MNIST and F-MNIST datasets compared to CNNs. However, the significant drawback is the high number of parameters required for operation. 
The introduction of DeepCaps in~\cite{deepcaps} utilizes a 3D convolution-based dynamic routing algorithm, showcasing improved performance on CIFAR10, SVHN, and Fashion MNIST datasets while reducing the parameter count by $68\%$. While several studies have explored FPGA-based accelerators for CNNs like AngelEye~\cite{angel}, the distinct computational flow inherent in the routing algorithm and the unique capsule structure pose challenges in applying these findings to capsule networks. In a recent study~\cite{fastcaps}, a two-step approach for deploying the CapsNet model on FPGA was proposed. Initially, the network is pruned using the LAKP approach, then optimized the dynamic routing algorithm to reduce complexity and enable parallel operations. In this study, we implemented a capsule network-based beamformer and compressed it using our LAKP-ML method. We simplified the dynamic routing algorithm and designed a specialized accelerator for the pruned and optimized CapsBeam model. Finally, we deployed the model on FPGA and evaluated its performance.
\subsubsection{Dynamic Routing Algorithm}
The dynamic routing algorithm between capsule layers serves as a fundamental building block in the capsule network. It maps the capsules (groups of neurons) in the lower layer to those in the upper layer by dynamically adjusting a parameter called the coupling coefficient. The three main operations involved in dynamic routing are softmax, fully connected transformation, and agreement, as described in Equations~\ref{eq1}-\ref{eq3}~\cite{caps}.
\begin{equation} \label{eq1}
    c_{ij} = \frac{\exp(b_{ij})}{\sum_k \exp(b_{ik})}   
\end{equation}
\begin{equation} \label{eq2}
     v_j = \text{squash} \left( \sum_i c_{ij} \hat{u}_{j|i} \right)
\end{equation}
\begin{equation} \label{eq3}
    b_{ij} \leftarrow b_{ij} + \hat{u}_{j|i} \cdot v_j
\end{equation}
Here, $\hat{u}_{ij}$ represents the input vector (also known as prediction vector for capsule $j$ in layer $l+1$ made by capsule $i$ in layer $l$), while $b_{ij}$ are the logits, initialized as zero. The final output vector of capsule $j$ is denoted as $v_j$. Squash is a nonlinear function that ensures the output vector remains within a unit norm. The process described in Equations~\ref{eq1}-\ref{eq3} is repeated for a specified number of routing iterations to refine the coupling coefficients and improve feature representation.

\subsection{Neural Network Pruning}
\par Network pruning enhances computational efficiency and reduces energy consumption by reducing parameter redundancy and minimizing memory references. Structured kernel pruning~\cite{kernal}, owing to its organized nature, is compatible with most neural network accelerators and embedded GPUs. Look-Ahead Kernel Pruning (LAKP)~\cite{fastcaps} leverages the sum of look-ahead scores of parameters and has demonstrated superior compression rates compared to magnitude-based kernel pruning~\cite{kernal}, especially in high-sparsity scenarios. The Look-Ahead Score for the $q^{th}$ kernel in the $p^{th}$ filter of the $i^{th}$ layer is calculated as the product of two terms: (1) the sum of the absolute values of the parameters in kernel $k(p,q)$, and (2) the summed absolute values of the parameters in its connected kernels from the $(i-1)^{th}$ and $(i+1)^{th}$ layers. Specifically, in the $(i-1)^{th}$ layer, the connected kernels belong to the $q^{th}$ filter, while in the $(i+1)^{th}$ layer, the connected kernels correspond to the $p^{th}$ channel across all filters. Finally, the kernels that attain the lowest look-ahead score will be pruned from the filter. In this work, we employ a multi-layer Look-Ahead Kernel Pruning method known as LAKP-ML, which examines the weights of more than one neighboring layers to compute the importance score of a kernel. 
\section{Proposed Methodology}
\par
\begin{figure}[!t]
\centering
\includegraphics[scale=0.39]{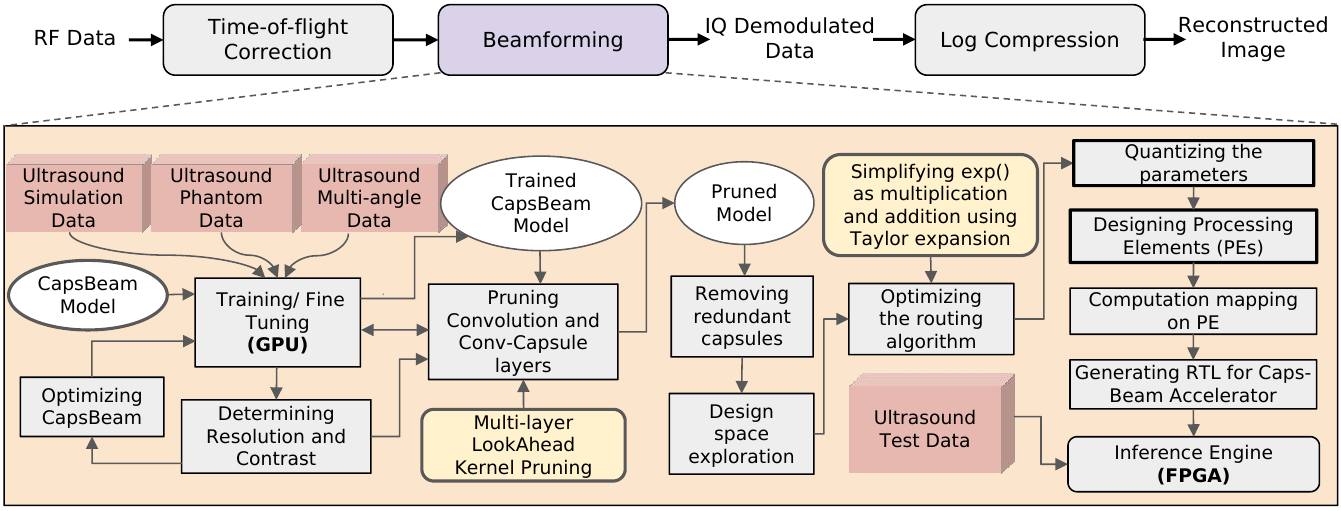}
\caption{An outline of the proposed methodology}
\label{fig:my_label-1}
\end{figure}

 In this section, we describe the implementation of the proposed CapsBeam model and its deployment on FPGA. An overview of the proposed methodology is shown in Fig.~\ref{fig:my_label-1}. The process starts with designing a beamformer model, CapsBeam, by refining the layers and its computation mapping to improve the throughput as well as the image quality. We utilized our structured multi-layer LookAhead Kernel Pruning (LAKP-ML) method and pruned the convolution layers in the network with contrast and resolution as the quality measure. The pruned model undergoes fine tuning to improve the image quality. Then we analysed the network and removed the redundant kernels and capsules. Furthermore, we conducted a comprehensive design space exploration and reduced the offchip memory transaction overhead as well as the hardware complexity by quantizing the model and optimizing the design flow. Finally, our pruned and optimized CapsBeam model is deployed on FPGA and evaluated the performance.  
\vspace{-2mm}
\subsection{CapsBeam Architecture}
\par
\begin{figure}[!t]
\centering
\includegraphics[scale=0.4]{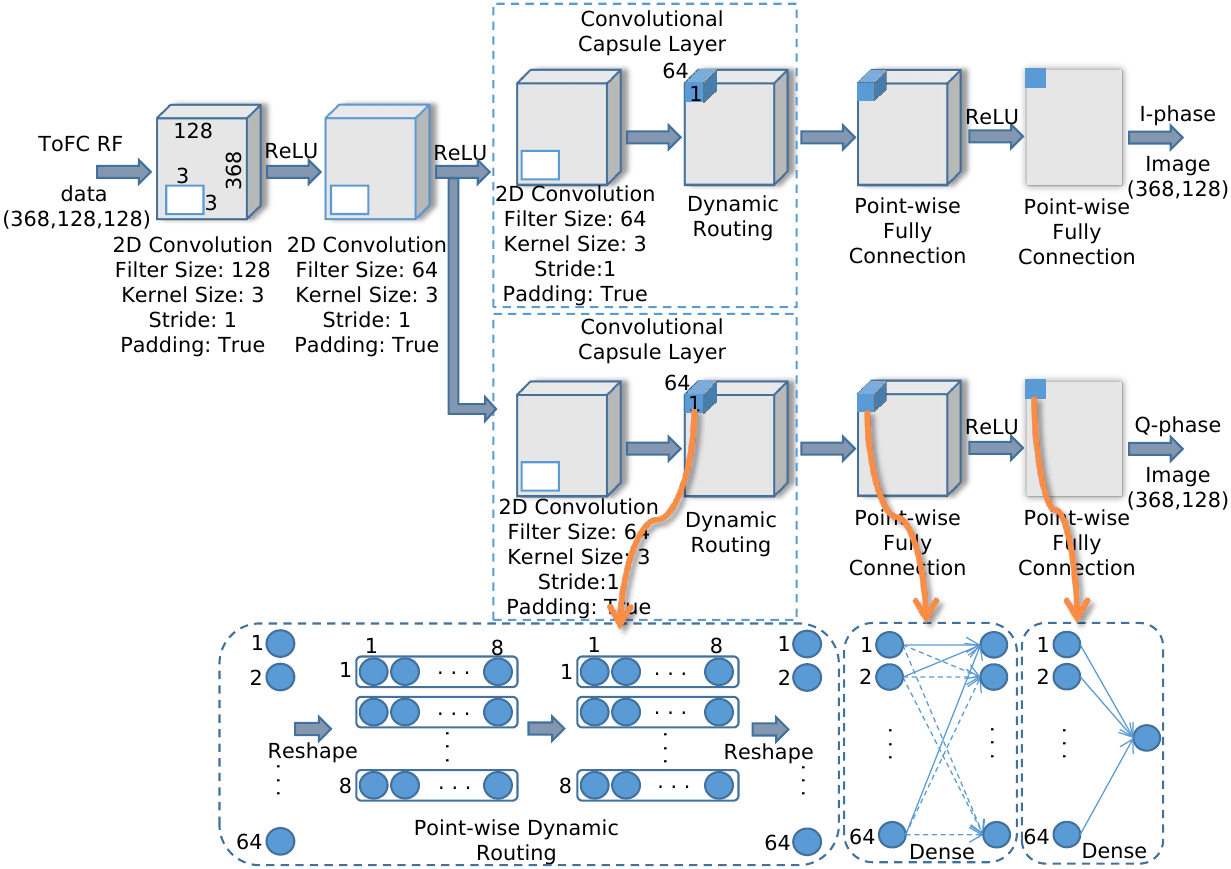}
\caption{The proposed CapsBeam architecture. The I-phase and Q-phase represent the real and Hilbert transform of the beamformed image, respectively. The convolutional capsule conducts 2D convolution, followed by pixel-wise dynamic routing, and finally, the pixel-wise fully connected layers to generate the I and Q values.}
\label{fig:my_label-2}
\end{figure}

 The architecture of the proposed CapsBeam model is depicted in the Fig.~\ref{fig:my_label-2}. It comprises of two convolution layers, two convolutional capsule layers, and four point-wise fully connected layers. The time-of-flight-corrected (ToFC) RF data is given as input to the model, which examines the signal values across the channels (elements in the transducer array). It then suppresses interference and noise from RF data and generates a high-quality envelope of the beamformed image.

  
\par The convolution layers within the model extract relevant features from the receptive field for each pixel in the beamformed image. Subsequently, the dynamic routing algorithm in the convolutional capsule layer, followed by the point-wise fully connected layers, maps these features to the envelop value (real and Hilbert transform values) for each pixel individually.
\par The ground truth for the model is the MVDR envelope of beamformed data. The covariance of the MVDR beamforming is computed with subarrays of length L=48 and temporal averaging over 2×(K=7)+1 samples. Similarly, the proposed model also considers the 7×7 receptive field (taking into account all four convolutions) for each pixel in the beamformed image and outputs performance similar to that of the MVDR beamformer with significantly less computational complexity.
\par
\begin{figure}[!t]
\centering
\includegraphics[scale=0.39]{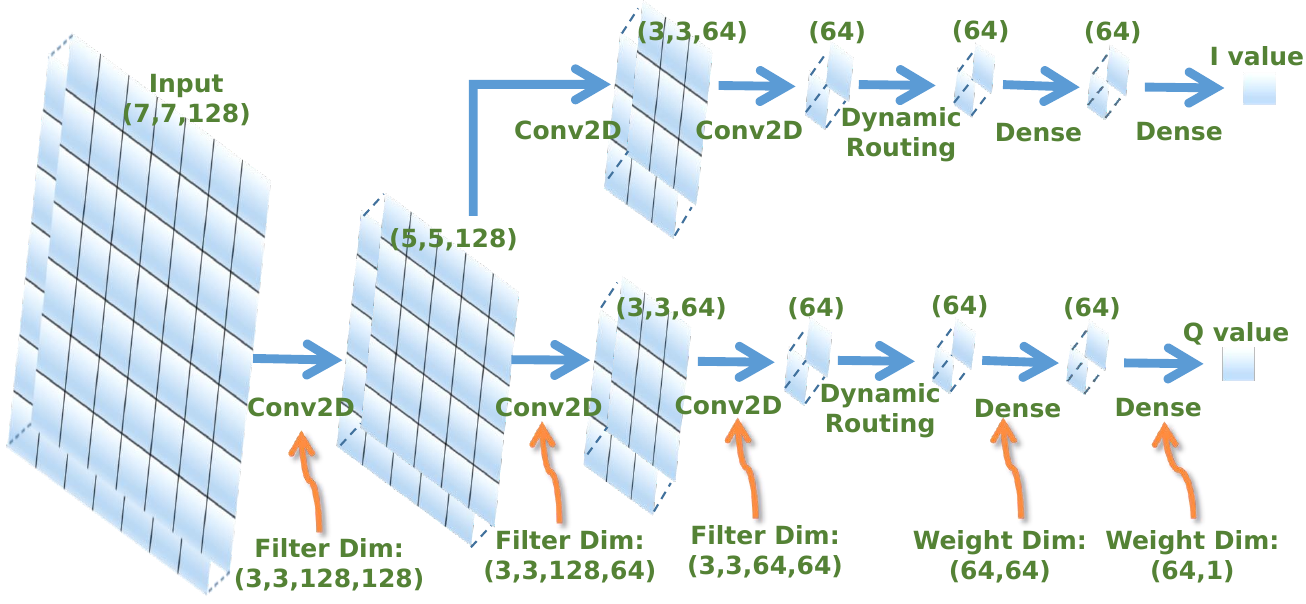}
\caption{Illustrates the flow of computing the envelope value (I and Q) using our CapsBeam model for each pixel}
\label{fig:my_label-3}
\end{figure}

 The depicted flow of computing the beamformed value using our CapsBeam model for each pixel value is illustrated in Fig.~\ref{fig:my_label-3}. Initially, the first two convolution operations and the subsequent two convolutions in the capsule layers construct a 7x7 receptive field. We leverage a modified version of the dynamic routing algorithm from the literature, tailored to process the features extracted from these receptive fields individually. Subsequently, the four fully connected layers compute both the real and Hilbert transformed output of the beamformed value for the specific pixel. Consequently, our model demonstrates rapid learning capabilities and exhibits performance similar to that of the target MVDR beamformer, all while maintaining significantly reduced computational complexity. To further refine the image quality, we fine-tuned the model using Coherent Plane-Wave Compounding (CPWC) beamformer output with 5 insonifications at angles of -0.86, -0.43, 0, 0.43, and 0.86 degrees.
\vspace{-2.5mm}
\subsection{Dataset}
\par The training dataset was collected in-house using the Verasonics Vantage 128 attached with an L11-5v transducer (Verasonics Inc., Redmond, WA)~\cite{data1,data2}. The acquisition procedure is detailed in~\cite{cnn,data1,data2}, and adhered to the guidelines set forth in the Helsinki Declaration of 1975, which were revised in 2000. Additionally, the model underwent fine tuning using multi-angled images from the CUBDL dataset with 5 transmission angles~\cite{cubdl}. The performance evaluation was conducted on datasets provided by the PICMUS~\cite{picmus}. 
\subsection{Parameter Setting}
\par The proposed approach involves various hyperparameters crucial for its execution. The learning rate was initially set at $10^{-4}$ and reduced to $10^{-6}$ for fine tuning. A batch size of 2 was chosen, and optimization utilized the Adam optimizer. The chosen loss function for this method is Mean Squared Error (MSE) employed on the I and Q components of the beamformed image before log compression.
\par The proposed CapsBeam model was implemented using the TensorFlow platform (version 2.40) and was trained and tested on a 12th Gen Intel\textsuperscript{\textregistered} Core\texttrademark\ i9-12900K × 24 system with an NVIDIA RTX A4500 GPU. The model comprises a total of 303,682 weights and has a computational complexity of approximately 28.79 GOP.

\subsection{Evaluation Metrics}
\par We employed four performance metrics to assess the resolution and contrast of the ultrasound reconstructed images. The axial and lateral resolutions were estimated using the full width half maximum (FWHM) measure at -6 dB of the pins/wires in the PICMUS dataset. Contrast was evaluated using three metrics: contrast ratio (CR)~\cite{cr}, contrast-to-noise ratio (CNR)~\cite{cnr}, and generalized contrast-to-noise ratio (gCNR)~\cite{gcnr}. 


\subsection{Multi-layer LookAhead Kernel Pruning (LAKP-ML)}
\par
\begin{figure}[!t]
    \centering
  
  \subfloat[\label{1a}]{%
        \includegraphics[scale=0.268]{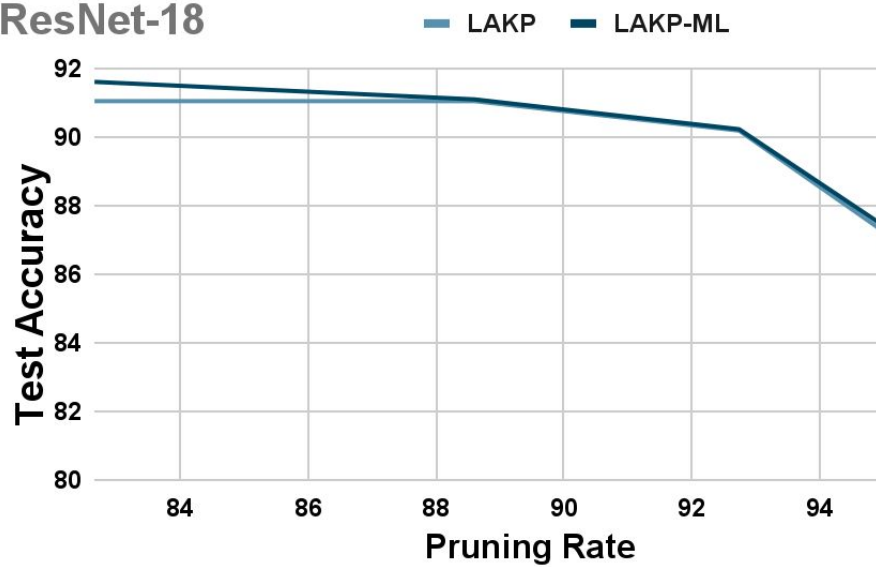}}
        \hspace{0.5mm}
  \subfloat[\label{1b}]{%
        \includegraphics[scale=0.25]{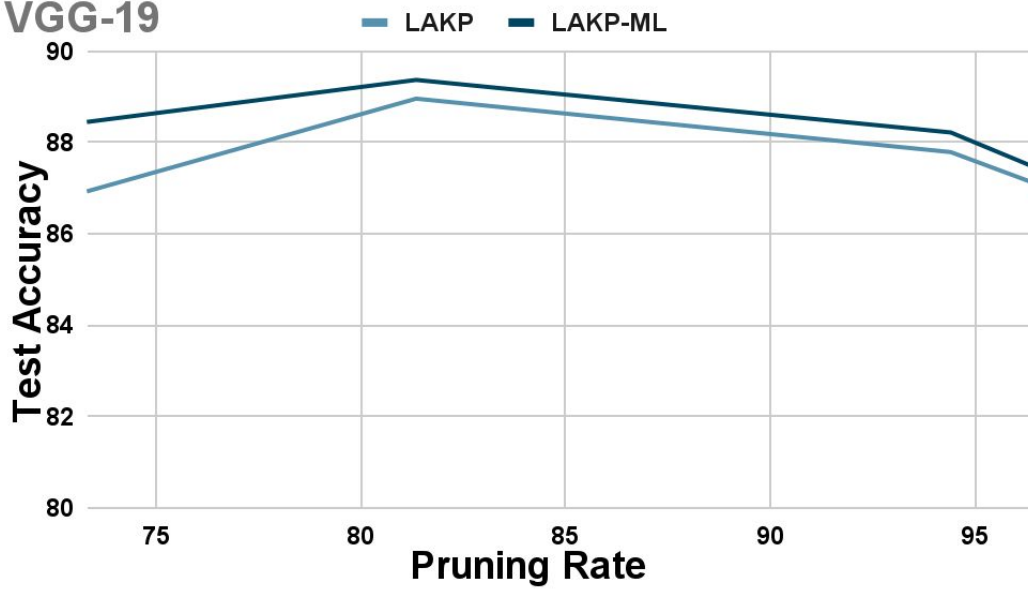}}
  \caption{Comparison of LAKP and LAKP-ML on (a) ResNet18 and (b) VGG-19 utilizing CIFAR-10 dataset and when r=2}
  \label{fig:my_label-4} 
\end{figure}

 Multi-layer LookAhead Kernel Pruning (LAKP-ML) is an extension of the structured LAKP method. LAKP examines the kernels in both the preceding and succeeding layers that are linked to the current kernel being pruned. LAKP showed good performance across standard models such as ResNet-18, VGG-19, CapsNet when evaluated on CIFAR-10, GTSRB, F-MNIST and MNIST datasets. LAKP-ML considers weights of multiple preceding and succeeding layers, rather than solely relying on the weights of direct neighboring layers, when pruning the current layer. The multi-layer pruning score for the kernel $k$ in the $i$-th layer can be computed as:
\begin{equation}
    P_i(k) = \left\|K_{i-r}[:]\right\| ... \left\|K_{i-1}[:]\right\|. \left\|k\right\| . \left\|K_{i+1}[:]\right\| ... \left\|K_{i+r}[:]\right\|
\end{equation}
where $k$ denotes the kernel in the $i$-th layer, $K_{i-1}[:]$ and $K_{i+1}[:]$ represent the subsets of kernels in the $(i-1)$-th and $(i+1)$-th layers, respectively, which are linked to the kernel $k$, and $r$ signifies the number of neighboring layers. Experiments conducted on ResNet-18 and VGG19 models demonstrated that LAKP-ML enhanced the accuracy of the pruned model compared to LAKP when evaluated on CIFAR-10 dataset, as shown in the Fig.~\ref{fig:my_label-4}. 
\par We utilized the LAKP-ML approach for pruning the kernels of the convolution layers in our CapsBeam model. Pruning the CapsBeam model differs from typical pruning methods used for classification models, as our focus is on preserving the resolution and contrast of the reconstructed images rather than optimizing classification accuracy. This iterative pruning process is repeated until the best-compressed model that maintains the original image quality is identified.

\subsection{Design Space Exploration}
\par Due to the large size of input, intermediate outputs and weights of each layer, it is not feasible to store all the parameters in the on-chip memory. Consequently, these parameters must be loaded online during inference, resulting in increased latency due to the overhead of external memory transactions. Partitioning the operations into tiles, storing a segment of the input and weights, and preparing a subset of outputs for subsequent layer operations is an effective approach to minimize latency. Additionally, these block operations can be executed in parallel by configuring processing elements and implementing an efficient data flow design. However, these optimizations are constrained by the resources availability on the target board.
\par In the case of the non-optimized CapsBeam model, the first layer input dimension is (368, 128, 128), and the filter dimension is (3, 3, 128, 128). If we consider on-chip BRAM for storing three rows of input and one set of filter weights (3, 3, 128) at a time, then the total external memory transaction overhead is calculated as $368 \times 3 \times 3 \times 128 \times 128 + 368 \times 128 \times 128= 60,293,120$. This occurs because the weight has to be repeated for each block of computation. However, this transaction becomes infeasible for a single transmission using the available DMA due to buffer overflow issues.
\par Our LAKP-ML method reduces the number of parameters to $15\%$ and enables the weights of each layer to be stored on on-chip memory, significantly reducing the external memory transaction overhead. During each layer operation, the complete weights are loaded once, while the input is partially loaded and then reloaded after the computation for that input is complete, ensuring that all parameters are loaded only once, thus reducing external memory transactions. Additionally, we have utilized two DMAs for transmitting input and weights in parallel, further reducing the latency of memory transactions.

\subsection{CapsBeam Accelerator}\label{sec-3G}
\par
\begin{figure}[!t]
\centering
\includegraphics[scale=0.4]{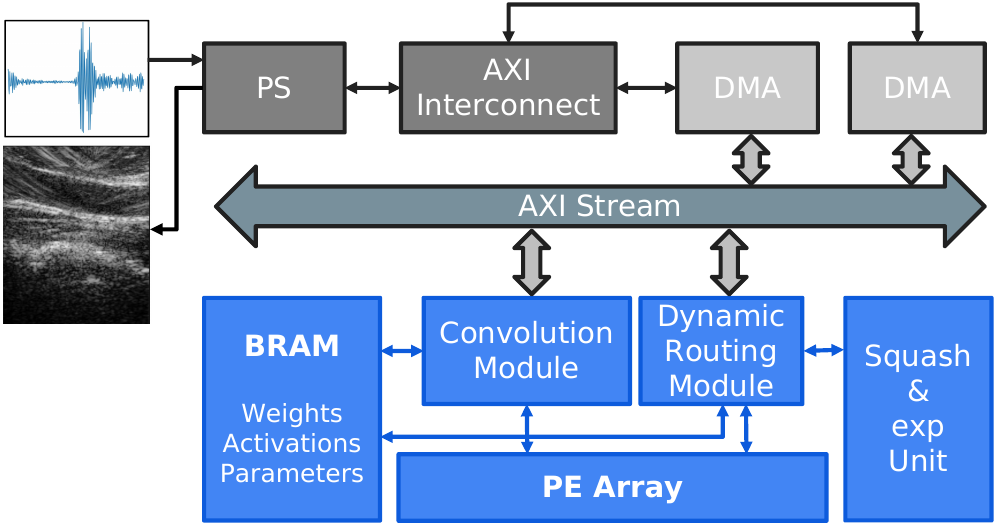}
\caption{The proposed CapsBeam accelerator architecture }
\label{fig:my_label-5}
\end{figure}

 The Fig.~\ref{fig:my_label-5} illustrates the accelerator architecture of the CapsBeam model designed for FPGA implementation. Our primary focus was on reducing external memory transactions and maximizing resource utilization. External off-chip memory is accessed through two DMAs: one for input RF data or intermediate output and another for model weights. This setup ensures that the model input and weights are transmitted and available simultaneously for the computation engine, allowing computation to commence without any unwanted delays. We employed a 16-bit vector bus for data transmission through the DMA, transmitting 4 data in a channel (64 Bytes) on each cycle from off-chip memory, thereby reducing the external data transmission overhead.
\par We implemented the convolution and dynamic routing as two separate modules due to their distinct computation flows. On-chip BRAM blocks were utilized for storing partial network weights, network inputs, and intermediate outputs. Additionally, there is another block dedicated to non-linear exponential and squash operations. All these blocks and data are coordinated through a controller module. We employed 16-bit fixed point quantization to minimize the resource consumption and utilized an array of processing elements (PEs) to enable parallel operations. The convolution accelerator architecture achieves a throughput of 30 GOPS, while the dynamic routing accelerator architecture achieves a throughput of 17.4 GOPS, despite the limited available hardware resources. Furthermore, the design flow of convolution and dynamic routing modules are described in Algorithm~\ref{algo-1} and Algorithm~\ref{algo-2} respectively.
\begin{algorithm}[!t]
\DontPrintSemicolon
  Stream in the weights, bias and index for the current layer utilizing the DMA-1\\
  Fill the first row of input BRAM block with zeros (padding)\\
  Stream in the first row of input activation for the current layer utilizing the DMA-2\\
  
  \For{row = 0 to $num\_rows$}    
    { 
    	\If{row != 0}
    	  {
    	  	Shift the data in second row of input BRAM to first row\\
            Shift the data in third row of input BRAM to second row
    	  }
    	\If{row == $num\_rows-1$}
    	  {
    	  	Fill the last (third) row of input BRAM with zeros (padding)
    	  } 
    	\Else
    	  {	
    	 	Stream in the next row of input activation utilizing the DMA-2
    	
    	  }	

        \For{ f = 0 to ($num\_filters$ / 4)}
        {
            Compute the convolution output for the current $row$ utilizing the PEs, producing partial output with 4 output channels with dimension ($num\_cols$, 4)
        }
        \If{ReLU = True}{ Apply ReLU to the output}
        Add bias to the output\\
        Stream out the output row with dimension ($num\_cols$, $num\_filters$) utilizing the DMA-2        
    }    
\caption{Convolution Module}
\label{algo-1}
\end{algorithm}

\begin{algorithm}[!t]
\DontPrintSemicolon
    
  \For{p = 0 to $num\_rows$ * $num\_cols$}    
    { 
    	Stream in the input activation with dimension (8, 8) utilizing the DMA-2\\
    	Initialize logits, $b$, with zeros\\
        Initialize coupling coefficients, $c$, with $softmax(b)$\\
        \For{ r = 0 to ($num\_routing$ / 4)}
        {
            \If{r $>$ 0}
              {
                Compute the $softmax(b)$ utilizing the exponential PEs 
              }
            Compute $s= input * c$ utilizing the PEs\\
            Compute $s= squash(s)$ utilizing the built-in $sqrt$ and the PEs\\
            Compute $b= input . s$ utilizing the PEs, where $.$ represents the dot product 
        }
        Stream out the output $s$ with dimension (8, 8) utilizing the DMA-2        
    }    
\caption{Dynamic Routing Module}
\label{algo-2}
\end{algorithm}
\subsubsection{Convolution architecture}
\par
\begin{figure}[!t]
\centering
\includegraphics[scale=0.4]{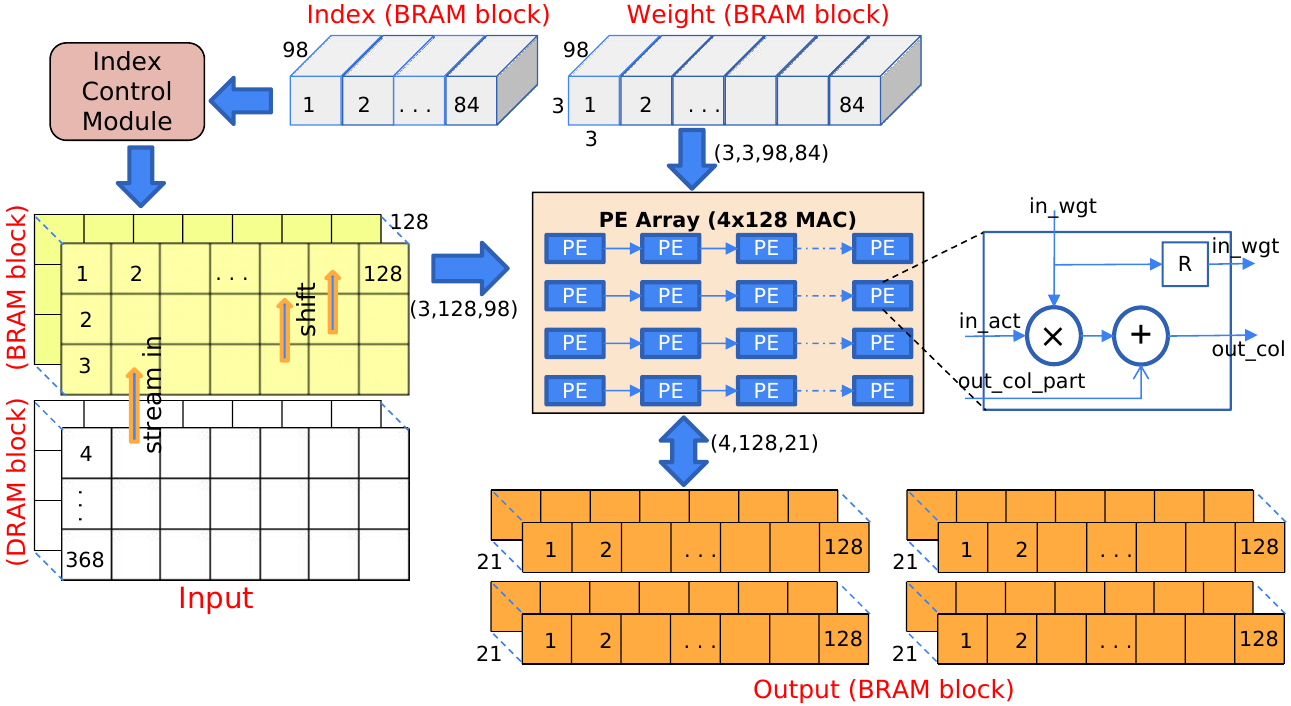}
\caption{The design flow of the convolution operation }
\label{fig:my_label-6}
\end{figure}

 We designed the convolution architecture with minimal external memory access. The micro-architecture of the convolution operation is shown in Fig.~\ref{fig:my_label-6}. For each convolution operation, the corresponding layer pruned weights along with its indices are stored to the BRAM block. The input is partially loaded into the BRAM block during each iteration. Initially, 3 rows of input (with dimensions of (3, 128, 128) for the first layer) are loaded into the BRAM. These are then processed with weights (with dimensions of (3, 3, 98, 84)), generating a single row of convolution output (with dimensions of (128, 84)). Subsequently, the next row of input values (with dimensions of (1, 128, 128)) is loaded, and computation is performed using the processing elements. This process continues until the complete convolution output is generated. 
\par The array of processing elements generates partial outputs with dimensions of (128, 4) during each cycle, where 128 represents the column dimension and 4 represents the output channel. These partial outputs are then reused for the next iteration. After processing all the weights, the computed single row of output with dimensions of (128, 84) is streamed to off-chip memory for the next layer operation. Meanwhile, the next input row values are stored in the BRAM block and the computation process continues.
\par We utilized an array of 4x128 processing elements (PEs) each performing single fixed point multiplication and addition operation. The input weights and input activations for each PE are multiplied together, and the result is accumulated with the partial output to generate a new partial output. 
\begin{figure}[!t]
\centering
\includegraphics[scale=0.45]{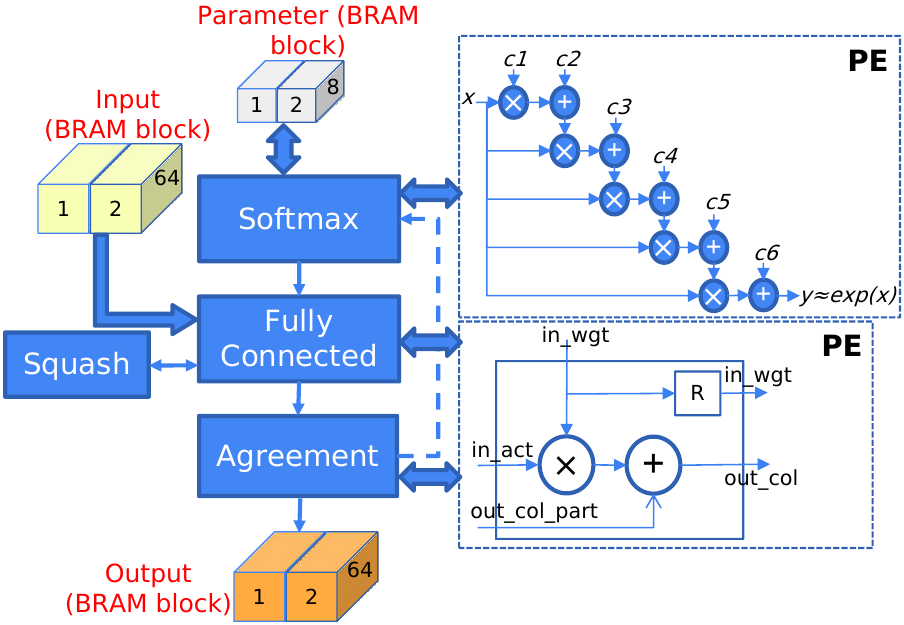}
\caption{The design flow of the dynamic routing operation}
\label{fig:my_label-7}
\end{figure}
\subsubsection{Dynamic Routing Architecture}
\par Due to the complex dataflow of dynamic routing algorithm, we designed it as a separate architecture module and performs computation with array processing elements and non-linear units. The micro-architecture of dynamic routing algorithm is demonstrated in the Fig.~\ref{fig:my_label-7}. It loads input values with dimension (2, 64) and generates dynamically routed outputs of the same dimension (2, 64) during each iteration. This process continues until all (368, 128, 64) values are generated. The Onchip BRAM block is utilized to store input, intermediate results, and dynamic routing parameters. The exponential operation within the Softmax() function is executed through five multiplications and five additions, leveraging the first five components of its Taylor expansion, as explained in~\cite{fastcaps}. The square root operation within the squash module is executed by invoking the built-in hls::sqrt() function present in the Vitis HLS tool. For the remaining dot product operations, array processing elements are utilized, with each conducting multiplication followed by addition. Additionally, partitioning of the BRAM blocks facilitates data availability for the computation engine.

\section{Experimental Results}
\par We evaluated our CapsBeam model using the publicly available in-silico, in-vitro, and in-vivo ultrasound datasets accessible through PICMUS. For the in-silico and in-vitro datasets, we assessed the performance of each beamformer in terms of contrast and resolution. Regarding the in-vivo data, we compared the performance of each beamformer with the CPWC beamformed output. CPWC is derived from 5 insonifications, while all other beamformed images are obtained from a single angle insonification of $0^{\circ}$.

\subsection{In-silico data}
\par
\begin{figure}[!t]
\centering
\includegraphics[scale=0.47]{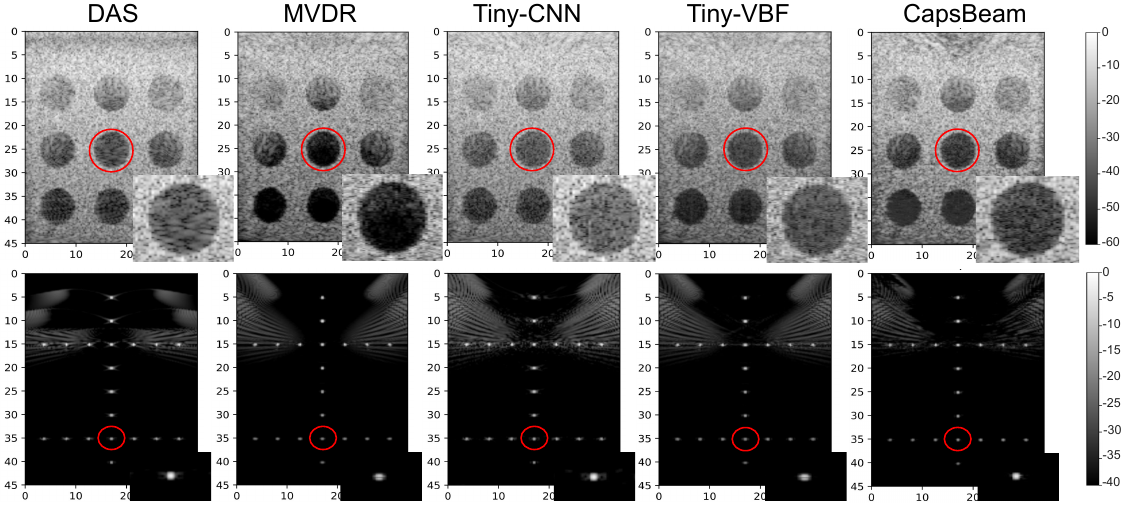}
\caption{Reconstrcuted images of standard DAS, MVDR, Tiny-CNN, Tiny-VBF and the proposed CapsBeam for in-silico contrast evaluation (top) and resolution-distortion dataset  (bottom)}
\label{fig:my_label-8}
\end{figure}

\begin{figure}[!t]
    \centering
  
  \subfloat[\label{1a}]{%
        \includegraphics[scale=0.24]{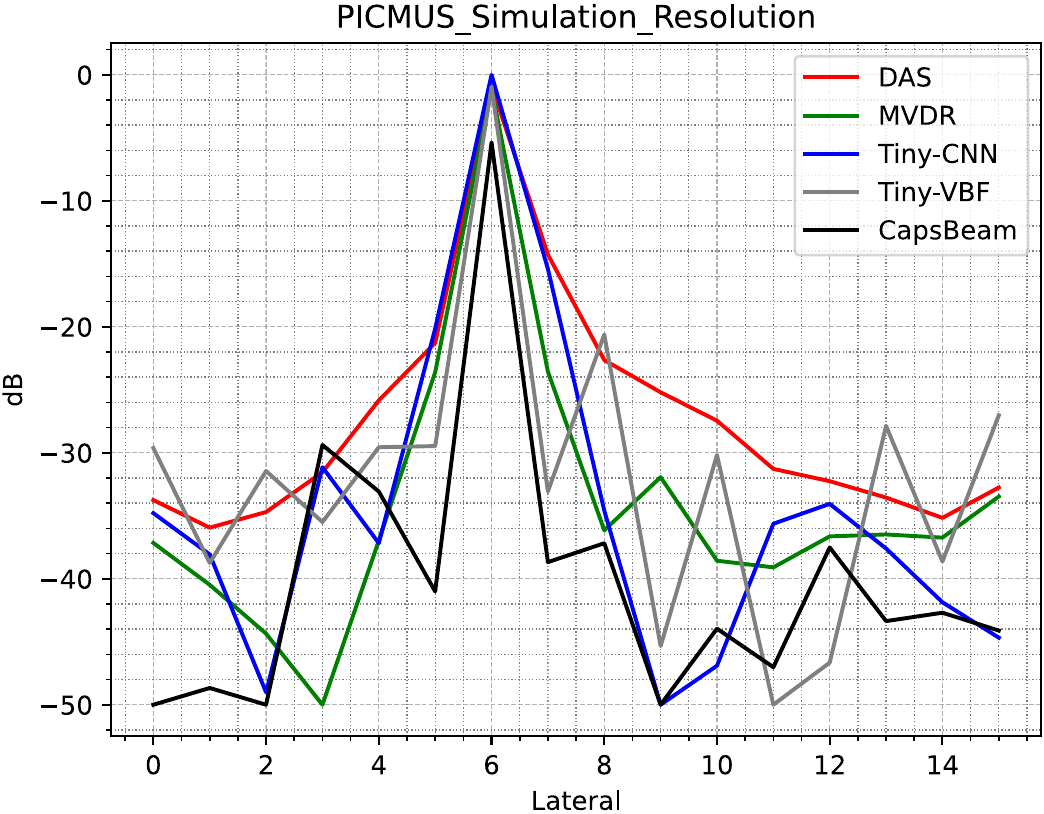}}
        \hspace{0.5mm}
  \subfloat[\label{1b}]{%
        \includegraphics[scale=0.24]{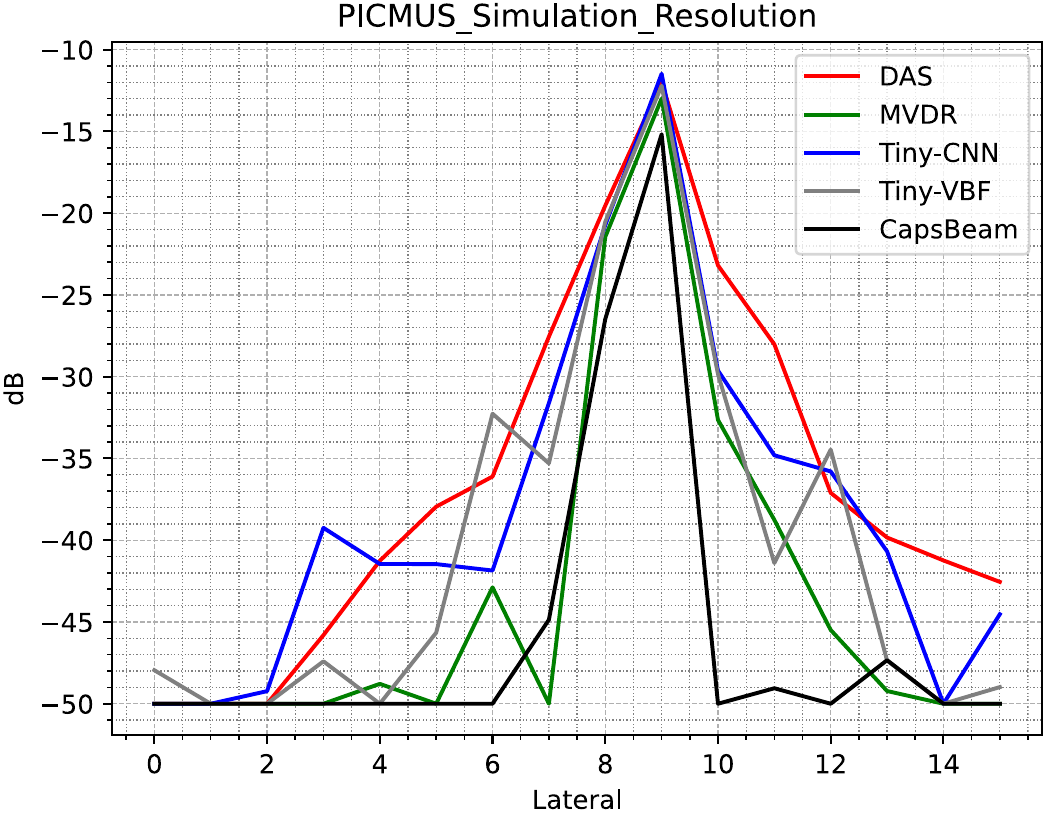}}
  \caption{Lateral profile at (a) 15.12 mm and  (b) 35.15 mm where x-axis represents the pixel position and y-axis represents the log compressed amplitude values}
  \label{fig:my_label-9} 
\end{figure}

 Fig.~\ref{fig:my_label-8} illustrates the results of different beamforming methods on the in-silico resolution and contrast data. The proposed CapsBeam yields high-contrast, high-resolution imaging that is qualitatively comparable to the MVDR target with significantly less clutter than DAS, Tiny-CNN and Tiny-VBF. In evaluating the cyst data, the performance of the state-of-the-art Tiny-CNN model is limited to that of the DAS. DAS and Tiny-CNN exhibit noise and other artifacts in the cyst region, which are significantly reduced in the case of CapsBeam. Furthermore, the quantitative results presented in the Table~\ref{table-1} indicate that the CR and GCNR are improved by $26\%$ and $3.61\%$ in the case of CapsBeam compared to DAS, respectively. Similarly, for point-scattered data, the performance of our CapsBeam is greatly improved compared to DAS and TinyCNN. The graphs presented in Fig.~\ref{fig:my_label-9} illustrate the lateral profile of pixel intensities along the lines of point scatters at depths of 15.12 and 35.15. It is evident from the results that the CapsBeam substantially reduces the mainlobe width and sidelobes compared to DAS and Tiny-CNN. Moreover, Table~\ref{table-2} shows the quantitative analysis of average axial and lateral resolution using different beamformers. We observe that our CapsBeam model improved the axial and lateral resolution by $13.6\%$ and $21.5\%$ as compared to DAS. 

\begin{table}[!t]
        \caption{Contrast metrics (mean) on in-silico and in-vitro data}
        \centering
        \begin{tabular}{p{0.18\linewidth}p{0.08\linewidth}p{0.08\linewidth}p{0.08\linewidth}p{0.08\linewidth}p{0.08\linewidth}p{0.08\linewidth}}
            \hline
            \textbf{Beamformer} & \multicolumn{3}{c}{\textbf{Simulation (dB)}} &  \multicolumn{3}{c}{\textbf{Phantom (dB)}} \\
            
            & CR & CNR & GCNR & CR & CNR & GCNR \\
            \hline
            DAS & 13.78 & \textbf{2.37} & 0.83 & 11.70 & 1.04 & 0.83 \\
            MVDR & \textbf{21.66} & 1.95 & 0.78 & 15.09 & \textbf{2.63} & 0.72 \\
            Tiny-CNN & 13.45 & 2.04 & 0.83 & 11.30 & 1.05 & 0.79 \\
            Tiny-VBF & 14.89 & 1.75 & 0.74 & 12.20 & 1.39 & 0.67 \\
            \textbf{CapsBeam} & \textbf{17.36} & 2.0 & \textbf{0.86} & \textbf{15.48} & \textbf{1.84} & \textbf{0.84} \\
            \hline 
        \end{tabular}
\label{table-1}
\end{table}

\begin{table}[!t]
        \caption{Quantitative measure interms of axial and lateral resolution on in-silico and in-vitro data}
        \centering
        \begin{tabular}{p{0.18\linewidth}p{0.08\linewidth}p{0.08\linewidth}p{0.08\linewidth}p{0.08\linewidth}}
            \hline
            \textbf{Beamformer} & \multicolumn{2}{c}{\textbf{Simulation (mm)}} &  \multicolumn{2}{c}{\textbf{Phantom (mm)}} \\
            
            & Axial & Lateral & Axial & Lateral  \\
            \hline
            DAS & 0.405 & 0.6 & 0.496 & 0.9 \\
            MVDR & \textbf{0.334} & \textbf{0.45} & 0.496 & \textbf{0.78} \\
            Tiny-CNN & 0.405 & 0.6 & 0.503 & 1.02 \\
            Tiny-VBF & 0.340 & \textbf{0.45} & 0.481 & \textbf{0.78} \\
            \textbf{CapsBeam} & 0.350 & 0.471 & \textbf{0.414} & 0.84 \\
            \hline 
        \end{tabular}
\label{table-2}
\end{table}

\subsection{In-vitro data}
\begin{figure}[!t]
\centering
\includegraphics[scale=0.47]{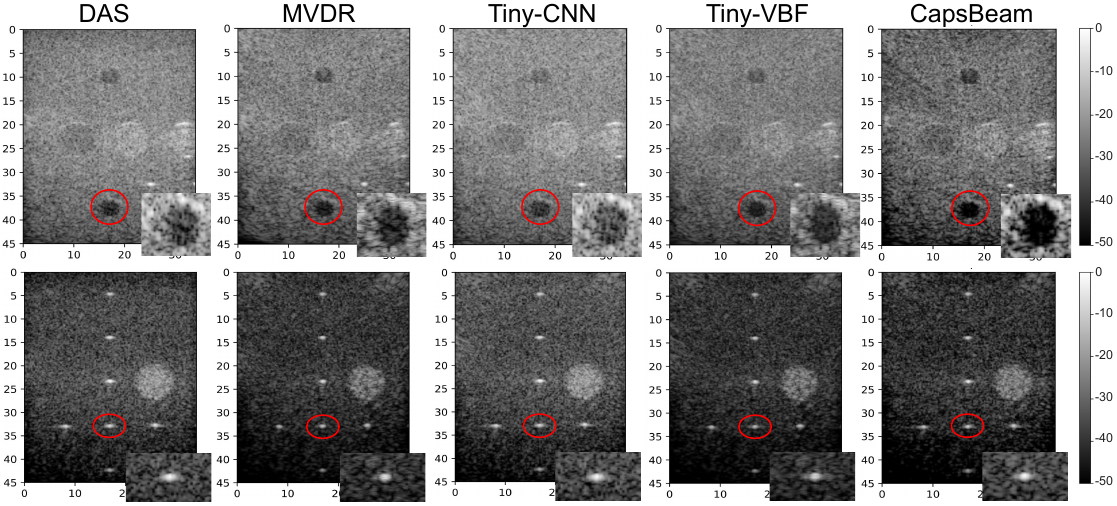}
\caption{Reconstrcuted images of standard DAS, MVDR, Tiny-CNN, Tiny-VBF and the proposed CapsBeam for in-vitro contrast evaluation (top) and resolution-distortion dataset (bottom) }
\label{fig:my_label-10}
\end{figure}

\begin{figure}[!t]
    \centering
  
  \subfloat[\label{1a}]{%
        \includegraphics[scale=0.24]{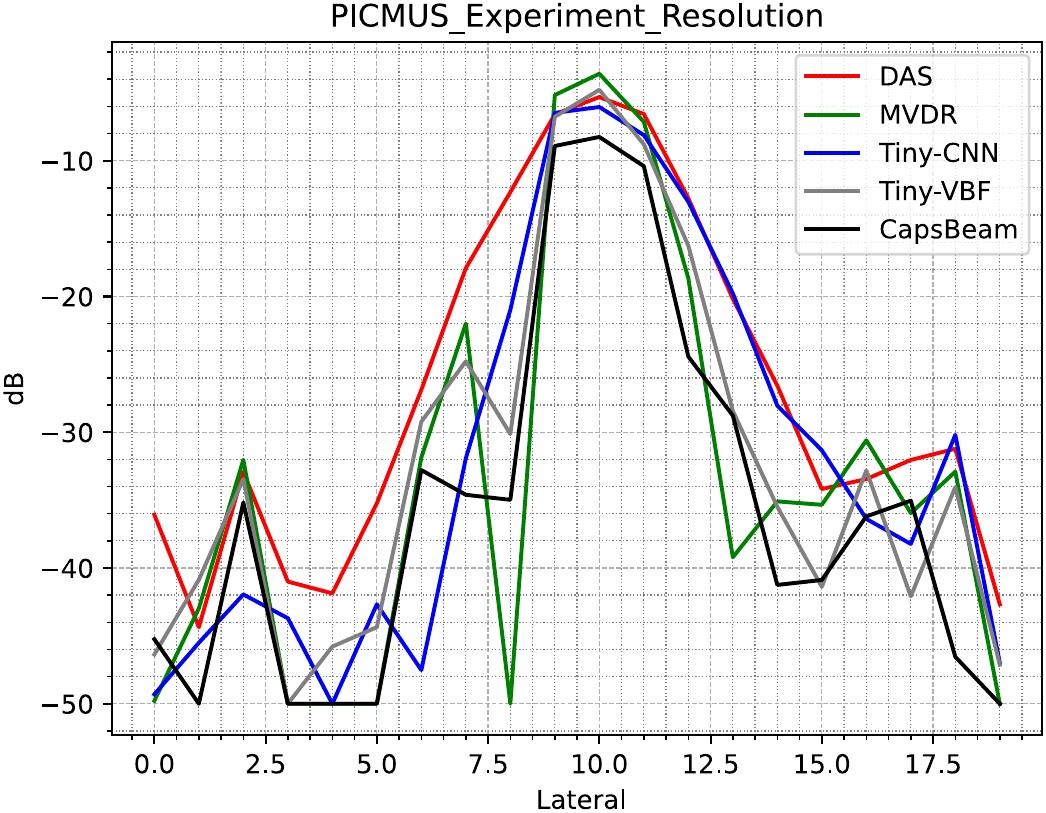}}
        \hspace{0.5mm}
  \subfloat[\label{1b}]{%
        \includegraphics[scale=0.24]{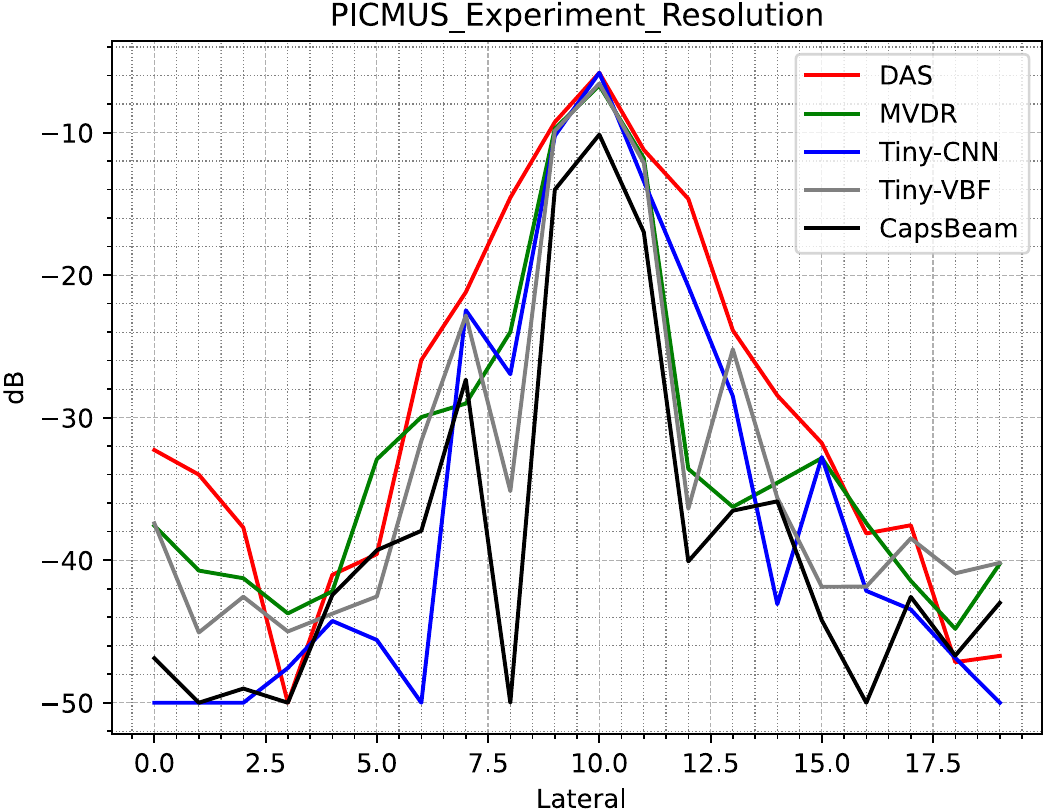}}
  \caption{Lateral profile at (a) 14.01 mm and  (b) 32.79 mm where x-axis represents the pixel position and y-axis represents the log compressed amplitude values}
  \label{fig:my_label-11} 
\end{figure}

We conducted experiments on in-vitro data and analyzed the contrast and resolution of DAS, MVDR, TinyCNN, Tiny-VBF, and our CapsBeam model. The reconstructed images are depicted in Fig.~\ref{fig:my_label-10}. Our CapsBeam model exhibits significantly improved contrast and resolution compared to other beamformers, as observed both qualitatively and quantitatively. The quantitative analysis is presented in the Table~\ref{table-1}, where the mean CR, CNR, and GCNR for the cyst targets demonstrate notable improvements with CapsBeam, showing improvements of $32.31\%$, $76.9\%$, and $1.2\%$ compared to DAS, respectively. Furthermore, CapsBeam enhances the axial and lateral resolution by $16.54\%$ and $6.7\%$, respectively, surpassing DAS, as demonstrated in the Table~\ref{table-2}. Additionally, the lateral profile of pixel intensities for contrast and resolution data is illustrated in Fig.~\ref{fig:my_label-11}. Our CapsBeam model effectively reduces overall artifacts and minimizes mainlobe width and sidelobes in the point scatter data.
\vspace{-2mm}

\subsection{In-vivo data}
\par
\begin{figure}[!t]
\centering
\includegraphics[scale=0.47]{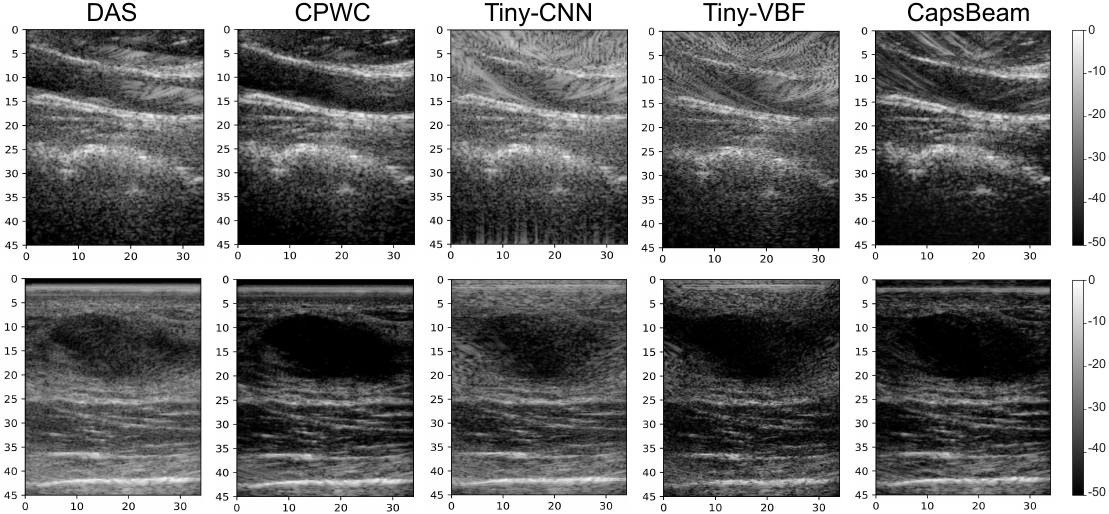}
\caption{Reconstrcuted images of standard DAS, MVDR, Tiny-CNN, Tiny-VBF and the proposed CapsBeam for in-vivo carotid artery (top) and breast mass dataset (bottom)}
\label{fig:my_label-12}
\end{figure}

 In addition to the in-silico and in-vitro data, we evaluated the performances of the beamformers on in-vivo data as well. Fig.~\ref{fig:my_label-12} displays the longitudinal view of the in-vivo carotid artery. We observe that the contrast in the artery region is significantly higher in the case of our CapsBeam model compared to DAS, Tiny-CNN, and Tiny-VBF. Additionally, clutter artifacts, such as white shades, are greatly reduced with our CapsBeam model. Similarly, for the CUBDL JHU data (bottom view of breast mass) shown in the figure, our CapsBeam model demonstrates superior performance compared to DAS, Tiny-CNN, and Tiny-VBF. We observe that contrast is maintained, and artifacts are reduced, resulting in images highly similar to CPWC beamformed output.

\vspace{-2mm}
\subsection{Evaluating LAKP-ML on CapsBeam}
\begin{figure}[!t]
\centering
\includegraphics[scale=0.5]{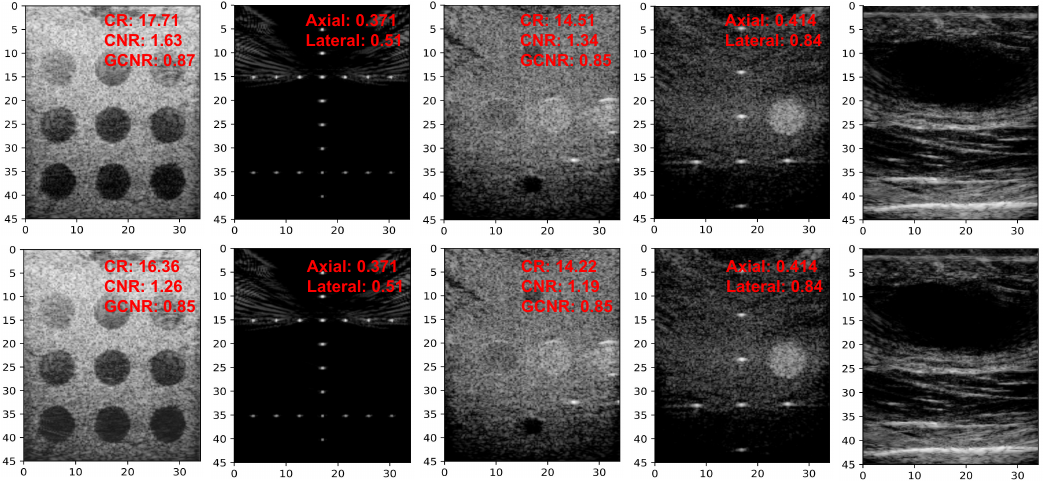}
\caption{Reconstrcuted images of pruned CapsBeam on CPU (top) and optimized CapsBeam on FPGA (bottom) for in-silico, in-vitro and in-vivo datasets}
\label{fig:my_label-13}
\end{figure}

Even though the CapsBeam model possesses a low number of parameters and computations, further model compression is necessary to reduce computation and facilitate deployment on resource-constrained edge devices. We pruned the CapsBeam model using our LAKP-ML method, achieving an effective compression ratio of 85$\%$ with minimal loss in image quality. Fig.~\ref{fig:my_label-13} depicts the beamformed output of the pruned CapsBeam model across different datasets, showcasing that the model maintains its reconstruction quality with 15$\%$ of its original weights. Additionally, from the quantitative measures, it is evident that while the axial and lateral resolution remain similar to the original model, there is a slight degradation in contrast.

\subsection{Evaluation on FPGA}
We deployed the 85$\%$ pruned model on the Xilinx ZCU104 evaluation board. The input to the accelerator is RF data with size of (368,128,128), and the final output from the beamformer is envelope data (I and Q data) with size (368,128). However, our accelerator design is compatible with any image size. We quantized the model using a 16-bit fixed-point format, which reduced computation and energy consumption. The reconstructed images on the FPGA are depicted in the Fig.~\ref{fig:my_label-13}. We observe that the quality of the images remains similar to that of the pruned CapsBeam model even after the optimizations. Additionally, the performance of Tiny-CNN and Tiny-VBF are evaluated on the FPGA. The convolutional architecture of the Tiny-CNN and non-optimized CapsBeam models follows a similar structure as demonstrated in Section~\ref{sec-3G}, apart from the absence of the index parameters. The Tiny-VBF accelerator is implemented based on the design presented in~\cite{tinyvbf}, with the additional optimizations demonstrated in Section~\ref{sec-3G}. Experimental results showed that Tiny-CNN, Tiny-VBF, and our non-optimized CapsBeam had computation latencies of 0.71, 1.49, and 1.35 seconds, respectively. Moreover, our optimized CapsBeam model significantly reduced the computation latency from 1.35 to 0.482 seconds, demonstrating a substantial improvement in efficiency.
Table~\ref{table-3} lists the amount of resources utilized by Tiny-CNN, Tiny-VBF, CapsBeam and the pruned and optimized CapsBeam model. The results demonstrate that, in the case of optimized CapsBeam, the combined effect of optimization and additional parallelism in the accelerator's computation led to a slight increase in LUT and DSP utilization while reducing BRAM usage. However, these enhancements significantly improved the overall throughput of image reconstruction, delivering higher performance and superior image quality compared to other beamformers.
\begin{table}[!t]
        \caption{Resource utilization of different tinyML beamformers on Xilinx ZCU104 evaluation board}
        \centering
        \begin{tabular}{p{0.15\linewidth}p{0.15\linewidth}p{0.15\linewidth}p{0.15\linewidth}p{0.15\linewidth}}
            \hline
            \textbf{Resources} & Tiny-CNN & Tiny-VBF & CapsBeam & Optimized CapsBeam \\
            
            \hline 
            \textbf{Slice LUTs} &72286      &134895    &107557     &163048 \\
                                &(31$\%$)   &(59$\%$) &(47$\%$)   &(71$\%$)\\
            \textbf{LUTRAM}     &1488       &24582     &2080       &22271\\
                                &(1$\%$)    &(24$\%$)  & (2$\%$)   &(22$\%$)\\
            \textbf{BRAM}       &278        &129      &283        &182.5 \\
                                &(89$\%$)   &(41$\%$) &(91$\%$)   &(59$\%$)\\
            \textbf{DSP48E}     &1038       &1298      &1490       &1363  \\
                                &(60$\%$)   &(75$\%$) &(86$\%$)   &(79$\%$)\\
            \textbf{Power (W)}  &5.611      &5.501    &5.96       &6.721 \\
            \hline
        \end{tabular}
\label{table-3}
\end{table}

\section{Discussion}
The primary objective in this work is to develop a hardware-optimized, lightweight beamformer for real-time ultrasound image reconstruction and implement a specialized accelerator, evaluating its feasibility and performance on a resource-constrained FPGA platform. State-of-the-art CNN-based approaches~\cite{unet,googlenet}, are primarily optimized for software-based applications and have a computational complexity exceeding 90 GOPs/Frame for a frame size of (368,128). These models are implemented on GPUs, specifically the Titan Xp and Nvidia GeForce GTX 1080Ti. However, their large parameter sizes and high computational demands make them impractical for resource-constrained edge devices. The Tiny-CNN~\cite{cnn} and Tiny-VBF~\cite{tinyvbf} introduced lightweight beamformers leveraging CNNs and vision transformers, respectively. Notably, Tiny-VBF was implemented on the Xilinx ZCU104 evaluation board, demonstrating feasibility for FPGA-based deployment. However, the reconstructed image quality is suboptimal, particularly for in-vivo datasets. In this study, we propose a capsule network-based lightweight beamformer that outperforms both standard DAS and existing tinyML beamformers~\cite{cnn,tinyvbf}. We implemented the beamformers, including DAS~\cite{das}, MVDR~\cite{mvdr}, CNN~\cite{googlenet}, Tiny-CNN~\cite{cnn}, and Tiny-VBF~\cite{tinyvbf}, and compared their frame rates against our CapsBeam model on the NVIDIA T4 Tensor Core and NVIDIA RTX A4500 GPUs, as shown in Table~\ref{table-4}. The results demonstrate that the non-optimized CapsBeam model achieved a significant throughput improvement, over $28\times$ compared to the MVDR beamformer and $2.5\times$ compared to the CNN beamformer, while preserving acceptable image quality. In contrast, Tiny-CNN and Tiny-VBF achieved approximately $1-2\times$ higher frame rates than the non-optimized CapsBeam model, but this came at the cost of reconstructed image quality, as discussed in Section IV.
\begin{table}[!t]
        \caption{Comparison of frame rate for different beamformers on GPU }
        \centering
        \begin{tabular}{p{0.35\linewidth}p{0.2\linewidth}p{0.2\linewidth}}
            \hline 
              & \multicolumn{2}{c}{\textbf{Frame Rate (FPS)}} \\
            \textbf{Beamformer}              & NVIDIA T4 Tensor Core&NVIDIA RTX A4500\\   
            \hline
            DAS~\cite{das}          & 714                   &1657\\
            MVDR~\cite{mvdr}        & 6                     &11\\
            CNN~\cite{googlenet}    & 67                    &150\\
            Tiny-CNN~\cite{cnn}     & 270                   &526\\
            Tiny-VBF~\cite{tinyvbf} & 179                   &757\\
            Non-optimized CapsBeam  & 172                   &380\\
            \hline
        \end{tabular}
\label{table-4}
\end{table}
\par We compressed the CapsBeam model to enhance computational efficiency using our multi-layer LookAhead Kernel Pruning (LAKP-ML), achieving an $85\%$ compression rate while preserving reconstruction quality. This parameter reduction significantly minimized off-chip data transmission and reduced computational complexity from 28.79 GOPs/Frame to approximately 6 GOPs/Frame for a frame size of (368,128). Additionally, we designed an FPGA-based accelerator architecture for the pruned CapsBeam model. Quantizing activations and parameters from 32-bit to 16-bit doubled the data transmission rate, while utilizing separate channels for weight and activation transmission further improved data throughput. We extended the exponential function simplification approach, initially evaluated for classification tasks in~\cite{fastcaps}, to beamforming. This optimization reduced the latency of dynamic operations from 0.11 seconds to 0.03 seconds. Furthermore, quantizing to fixed-point arithmetic, increasing parallelism, and simplifying the exponential operation significantly optimized computational efficiency, reducing the overall latency from 1.35 seconds (non-optimized) to 0.482 seconds when implemented on the Xilinx ZCU104 evaluation board. To benchmark performance, we implemented the state-of-the-art CNN (with an accelerator architecture similar to Tiny-CNN), Tiny-CNN and Tiny-VBF on the Xilinx ZCU104 evaluation board and evaluated their frame rates. Experimental results showed that our optimized CapsBeam achieves 2.07 FPS, surpassing state-of-the-art models, including CNN (0.015 FPS), Tiny-CNN (1.41 FPS), and Tiny-VBF (1.48 FPS). This highlights CapsBeam's efficiency in both throughput and image quality compared to other beamformers.
\par In our comprehensive evaluation of different beamformers using resolution (FWHM) and contrast (CR, CNR, GCNR) quality metrics, the proposed CapsBeam model consistently outperformed DAS, Tiny-CNN, and Tiny-VBF, particularly on in-vivo data. While CapsBeam demonstrated superior resolution compared to DAS and Tiny-CNN, it showed a slight lateral resolution degradation of approximately $7.6\%$ on in-silico and in-vitro data compared to Tiny-VBF, as shown in Table~\ref{table-2}. This indicates that Tiny-VBF achieves better resolution in smooth regions where artifacts in the retrieved signal are minimal. Notably, the in-silico evaluations involve computational simulations, while the in-vitro evaluations use phantoms containing wires or cysts at fixed locations, both specifically designed to assess resolution and contrast. However, in the case of in-vivo data collected from living tissue, where various artifacts are present, CapsBeam effectively minimizes these distortions, as illustrated in Fig.~\ref{fig:my_label-13}, outperforming Tiny-VBF, Tiny-CNN, and DAS. This highlights CapsBeam's superior generalization capability in real-world imaging scenarios.


\section{Conclusion}
In this study, we developed a novel capsule network-based adaptive beamformer for ultrasound plane wave imaging. Experiments conducted on in-silico, in-vitro and in-vivo data showed that the CapsBeam model improved resolution and contrast compared to the standard DAS beamformer. To reduce the computational load of CapsBeam, we pruned the model using our LAKP-ML approach and compressed model parameters by $85\%$. Additionally, we reduced the hardware complexity of the model through quantization and simplification of non-linear operations. Finally, we designed a specialized accelerator architecture for our CapsBeam model and evaluated its performance on the Xilinx ZCU104 board. Our accelerator achieved a throughput of 30 GOPS for convolution and 17.4 GOPS for dynamic routing operation.

\balance

\begin{IEEEbiography}[{\includegraphics[width=1in,height=1.25in, clip]{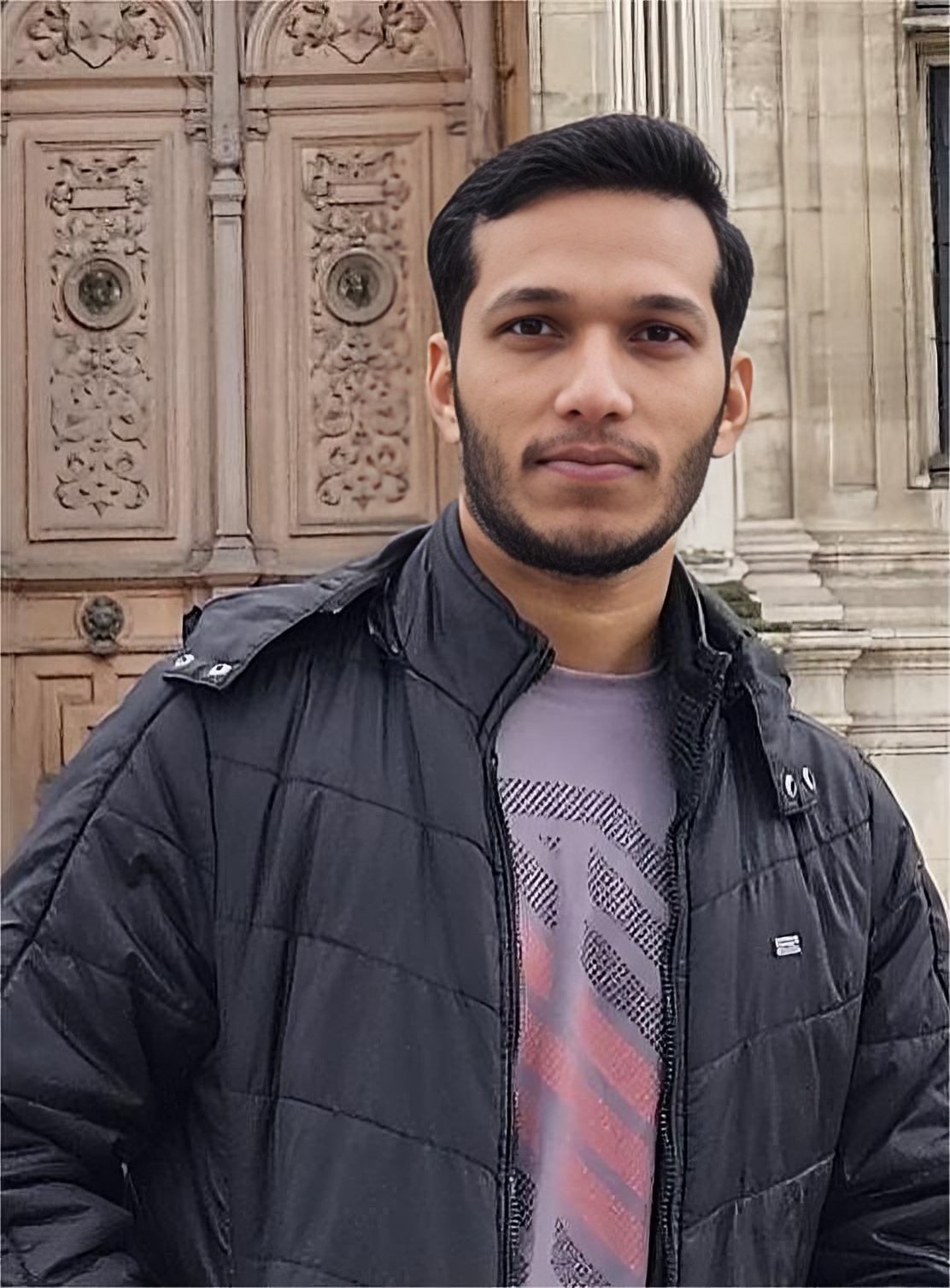}}]{Abdul Rahoof} received the B.Tech. degree in Computer Science and Engineering from Lal Bahadur Shastri College of Engineering, Kasaragod, Kerala, India, in 2017, and the M.Tech. degree in Information Security from the College of Engineering, Trivandrum (CET), Kerala, in 2019. He is currently pursuing the Ph.D. degree at the Indian Institute of Technology (IIT) Palakkad, Kerala. His current research focuses on the application of deep learning techniques and hardware acceleration for complex algorithms, particularly in the domain of medical imaging. His broader research interests include hardware optimization and machine learning.
\end{IEEEbiography}

\begin{IEEEbiography}[{\includegraphics[width=1in,height=1.25in, clip]{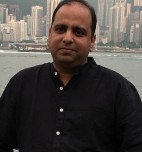}}]{Vivek Chaturvedi} (Member, IEEE) is currently working as Associate Professor in the Department of Computer Science and Engineering at Indian Institute of Technology, Palakkad, India. Previously, he was a Research Scientist in the School of computer science and engineering at Nanyang Technological University, Singapore. He received his M.S. from Syracuse University, NY and Ph.D. from Florida International University, Miami. He has also worked in Sun Microsystems as a Student intern. Dr. Chaturvedi's current research interest includes power, thermal and lifetime reliability optimization in multi/many core processors including both 2D and 3D architectures. He is also actively working on Hardware acceleration of complex algorithms and Adversarial Machine Learning. He has served as reviewer and TPC for several IEEE/ACM Journals and Conferences.
\end{IEEEbiography}

\begin{IEEEbiography}[{\includegraphics[width=1in,height=1.25in, clip]{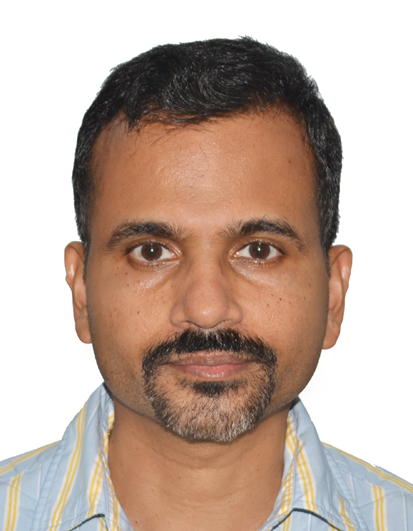}}]{Mahesh Raveendranatha Panicker} (Senior Member, IEEE) received his B. Tech. degree in Electrical and Electronics Engineering from Rajiv Gandhi Institute of Technology, India in 2003 and Ph.D. degree from the School of Computer Engineering, Nanyang Technological University (NTU), Singapore in 2009. Post a two-year stint as postdoctoral fellow at NTU, Mahesh transitioned to industry, working as a Lead Research Engineer at GE Global Research Centre, Bangalore, from 2010 to 2017. Subsequently, he served as Senior Chief Engineer at Samsung Research Institute, Bangalore, until 2018. He entered academia as an Assistant Professor of Electrical Engineering at IIT Palakkad in 2018, and later became an Associate Professor in 2021. Since December 2023, he has been an Associate Professor at the Singapore Institute of Technology's Infocomm Technology Cluster. His research interests include signal/image processing, machine learning, and embedded and accelerated computing, with applications in diagnostic ultrasound, biomedical systems, and industrial prognostics.
\end{IEEEbiography}

\begin{IEEEbiography}[{\includegraphics[width=1in,height=1.25in,clip, keepaspectratio]{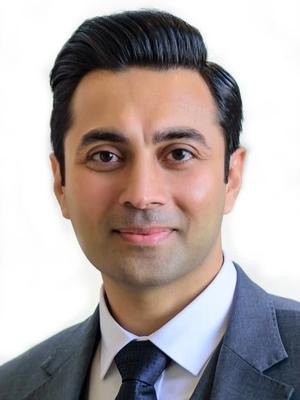}}]{Muhammed Shafique} (Senior Member,
IEEE) received the Ph.D. degree in computer
science from the Karlsruhe Institute of Technology
(KIT), Germany, in 2011.

He established and led a highly recognized research group at KIT for several years
and conducted impactful collaborative research
and development activities across the globe.
In October 2016, he joined the Institute of Computer Engineering, Faculty of Informatics, Technische Universität Wien (TU Wien), Vienna, Austria, as a Full Professor of
computer architecture and robust, and energy-efficient technologies. Since
September 2020, he has been with New York University (NYU), where he is
currently a Full Professor and the Director of the eBrain Laboratory, NYU,
Abu Dhabi, United Arab Emirates, and a Global Network Professor with
the Tandon School of Engineering, NYU, New York City, USA. He is also
a Co-PI/Investigator in multiple NYUAD Centers, including the Center of
Artificial Intelligence and Robotics (CAIR), the Center of Cyber Security
(CCS), Center for Interacting Urban Networks (CITIES), and the Center
for Quantum and Topological Systems (CQTS). He holds one U.S. patent.
He has (co)authored six books, more than ten book chapters, more than
350 papers in premier journals and conferences, and more than 100 archive
articles. His research interests include AI and machine learning hardware
and system-level design, brain-inspired computing, machine learning security and privacy, quantum machine learning, cognitive autonomous systems, wearable healthcare, energy-efficient systems, robust computing, hardware security, emerging technologies, FPGAs, MPSoCs, embedded systems,
cross-layer analysis, modeling, design, and optimization of computing and
memory systems. The researched technologies and tools are deployed in
application use cases from the Internet of Things (IoT), smart cyber-physical
systems (CPS), and ICT for development (ICT4D) domains. He is a Senior
Member of the IEEE Signal Processing Society (SPS) and a member of
ACM, SIGARCH, SIGDA, SIGBED, and HIPEAC. He received the 2015
ACM/SIGDA Outstanding New Faculty Award, the AI 2000 Chip Technology Most Influential Scholar Award, in 2020 and 2022, and the ASPIRE
AARE Research Excellence Award, in 2021. He received six gold medals and
several best paper awards and nominations at prestigious conferences. He has
served as the PC chair, the general chair, the track chair, and a PC member for
several prestigious IEEE/ACM conferences. He has given several keynotes,
invited talks, and tutorials as well as organized many special sessions at
premier venues.
\end{IEEEbiography}

\vfill

\end{document}